\documentclass[letterpaper,10pt,conference]{ieeeconf}

\IEEEoverridecommandlockouts
\usepackage{amsmath}
\usepackage{amssymb}
\usepackage[nocompress,nobreak]{cite}
\usepackage{graphicx}
\usepackage{float}
\usepackage{stfloats}
\usepackage[ruled,linesnumbered]{algorithm2e}
\SetAlgoLined
\SetArgSty{textnormal}
\usepackage[caption=false,font=footnotesize]{subfig}
\SetKwInput{KwIn}{Input}
\SetKwInput{KwOut}{Output}
\usepackage{placeins}
\usepackage{needspace}
\usepackage[bottom]{footmisc}
\usepackage{url}
\usepackage{booktabs}
\usepackage{xcolor}

\usepackage{microtype}
\makeatletter
\let\@IEEEoriginalmakecaption\@makecaption
\long\def\@makecaption#1#2{%
  \ifx\@captype\@IEEEtablestring
    \vskip 2pt
    {\centering\footnotesize #1\\\scshape #2\par}%
    \vskip 5pt
  \else
    \@IEEEoriginalmakecaption{#1}{#2}%
  \fi}
\makeatother

\title{\LARGE \bf
Mode Recommendation and Dynamic Routing Guidance for \\ Improved Trip Quality in Mixed Traffic Networks}

\author{Jin Zhang, Andreas A. Malikopoulos, and Ting Bai%
\thanks{This research was supported in part by the Research Start-UP Fund of Shanghai Jiao Tong University under Grant WL220203203, and in part by the National Natural Science Foundation of China under Grant 62603450.}%
\thanks{J. Zhang and T. Bai are with the School of Automation and Intelligent Sensing, Shanghai Jiao Tong University, Shanghai, China.
E-mails: \texttt{\{planckjin, tingbai\}@sjtu.edu.cn}}
\thanks{Andreas A. Malikopoulos is with the Applied Mathematics, Systems Engineering, Robotics, Mechanical Engineering, Electrical and Computer Engineering, and Civil \& Environmental Engineering, Cornell University, Ithaca, NY, USA. Email: \texttt{amaliko@cornell.edu}}}

\begin{document}
\bstctlcite{BSTcontrol:fullauthors}

\maketitle
\thispagestyle{empty}
\pagestyle{empty}
\flushbottom

\begin{abstract}
Urban mixed traffic networks provide diverse mobility options, yet conventional routing guidance with fixed travel modes leaves the benefits of coordinating travel modes and route planning largely unexploited. This paper proposes an integrated mode recommendation and dynamic routing guidance framework for urban mixed traffic networks comprising private vehicles, autonomous vehicles, and buses. We formulate a dynamic trip index (DTX) that incorporates mode-specific travel time, monetary cost, and convenience, enabling consistent evaluation of trip quality across travel modes. Using this measure, we develop a mode recommendation method that selects feasible autonomous vehicle or bus services based on predicted DTX for travelers without private vehicles, while preserving service capacity for travelers with known demand. We integrate this method with dynamic routing guidance for private and autonomous vehicles to improve the average trip quality of all travelers. Simulation studies involving $1{,}500$ travelers on a real-world road network in Shanghai, China, show that the proposed framework increases the mean realized DTX by $13.01\%$ compared with dynamic routing guidance alone and by $19.68\%$ compared with fixed travel modes with preplanned routes, highlighting the value of mode flexibility in improving trip quality beyond route adaptation.
\end{abstract}

\section{Introduction}
Urban mixed traffic networks support a broad range of urban activities, including commuting, employment, shopping, and social interactions~\cite{10932667}. These networks integrate diverse mobility options, such as private vehicles, autonomous mobility services, and public transit~\cite{zardini2022analysis,zardini2023codesign}. In reality, travelers' choices among these options are influenced by various factors, such as income, time constraints, and preferences for convenience~\cite{malokin2019}, resulting in increasingly diverse travel needs. Despite these differences in mobility preferences, all travelers share the underlying transportation infrastructure and compete for limited road capacity. Traffic conditions and congestion thus affect the travel experience across different modes, making \textit{trip quality} a key consideration for travelers throughout their daily journeys~\cite{bai2025routing}. In this context, trip quality depends not only on the selected travel mode but also on the route taken by the vehicle and the traffic conditions encountered along the route.

Existing routing guidance approaches improve travel performance by adapting vehicle routes to changing traffic conditions~\cite{10379532,11615763} and coordination targets~\cite{10209062,11312531}, but typically assume that travel modes are fixed. This assumption overlooks the potential benefits of jointly optimizing mode recommendation and route plans~\cite{lai2023travel}. For travelers without private vehicles, choosing among available mobility options offers an additional way to improve trip quality. However, mode choice, route selection, and traffic conditions are closely interrelated. Assigning travelers to different modes redistributes traffic flows across road segments, thereby affecting congestion and trip quality. In addition, limited service capacity may render certain mobility options unavailable to some travelers due to other travelers’ mode choices. Motivated by these considerations, this paper develops an integrated framework that coordinates feasible mode recommendations with dynamic routing guidance to improve the average trip quality of travelers across the network.

\subsection{Related Work}
Research related to this work falls into two main streams: routing
guidance and travel mode recommendation. The former acts on vehicle routes
during travel, whereas the latter acts on travelers' mobility options before
departure. Existing routing guidance
approaches use traffic information to select or update routes to reduce travel time and alleviate congestion~\cite{pan2013proactive,ho2023collaborative}.
Other approaches incorporate energy consumption costs into route
optimization~\cite{houshmand2022combined}, or integrate travel time, monetary
cost, and convenience to support equity-oriented routing
guidance~\cite{bai2025routing}. Travel mode recommendation methods draw on travelers’ historical choices, personal characteristics, and spatial and temporal context to recommend a single mode or a combination of modes~\cite{liu2021exploring,yan2024improving}. Preference-based trip planners further evaluate public transit and shared mobility options using travel time, monetary cost, and user preferences, while accounting for service availability~\cite{zhang2025dynamic}. {Intermodal fleet studies use network flow models to coordinate autonomous vehicle fleets with public transit at the network level~\cite{salazar2020intermodal,wollenstein2022}.}

However, existing studies generally address route adaptation or mode selection separately, with limited attention to the coupling between the two. Routing guidance methods~\cite{pan2013proactive,ho2023collaborative,houshmand2022combined} take each vehicle's travel mode as given and improve trip quality only through route changes within that mode. Mode recommendation
approaches~\cite{liu2021exploring,yan2024improving,zhang2025dynamic} evaluate candidate modes using travel attributes estimated before departure,
but rarely model how the recommended trips alter traffic conditions. {These intermodal studies capture this coupling but optimize aggregate flows under constant travel demand, without recommending modes to individual travelers or updating vehicle routes during travel.} In addition, few studies account for limited service capacity, under which assigning a vehicle or bus seat to one traveler may reduce the options available to other travelers. Treating mode recommendation and routing as separate decisions may overlook how mode recommendations influence subsequent route planning and how route adaptation changes the relative benefits of available modes.

\subsection{Contributions}
To address the challenges of coordinating feasible mode recommendations with dynamic routing, this paper proposes an integrated framework to improve average trip quality in mixed traffic networks. Different from our previous work in~\cite{bai2025routing}, which assumes fixed travel modes, the proposed framework combines mode recommendations for individual travelers, capacity reservations, and dynamic routing guidance. The main contributions are twofold: (i) We extend the dynamic trip index (DTX) introduced in~\cite{bai2025routing} to evaluate bus trips and develop a mode recommendation method based on predicted DTX values. The method recommends modes to travelers in sequence and reserves autonomous vehicles and bus seats for travelers who have not yet received recommendations. (ii) We integrate the mode recommendation method with {dynamic routing guidance that updates vehicle routes during travel to minimize the predicted remaining travel time.} Simulation studies involving 1,500 travelers on a real-world road network in Shanghai demonstrate substantial improvements in mean realized DTX compared to baselines with fixed travel modes and preplanned routes and dynamic routes alone.

\FloatBarrier
\raggedbottom
\setlength{\abovedisplayskip}{4pt plus 2pt minus 1pt}
\setlength{\belowdisplayskip}{4pt plus 2pt minus 1pt}
\setlength{\abovedisplayshortskip}{2pt plus 1pt minus 0.5pt}
\setlength{\belowdisplayshortskip}{4pt plus 2pt minus 1pt}
\setlength{\jot}{2pt}
\setlength{\intextsep}{6pt plus 2pt minus 1pt}

\section{Problem Formulation}
\label{sec:problem_formulation}

\pdfglyphtounicode{parenleftbigg}{0028}
\pdfglyphtounicode{parenrightbigg}{0029}

\subsection{Road Network}
\label{sec:road_network}
We model the road network as a directed graph $\mathcal{G}\!=\!(\mathcal{V},\mathcal{E})$, where $\mathcal{V}$ and $\mathcal{E}\subseteq\mathcal{V}\times\mathcal{V}$ denote the node set and directed edge set, respectively. Each node $v\in\mathcal{V}$ represents a road junction or a bus stop, and each edge $(v,v')\in\mathcal{E}$ represents a directed road segment connecting two nodes. Road segments are split at bus stops so that each bus stop is explicitly represented as a node. The weight
$\tau(v,v')\in\mathbb{R}_{+}$ of each edge denotes the travel time to traverse the edge,
where $\mathbb{R}_{+}$ is the set of positive real numbers. We denote by
$\tau^{0}(v,v')$ the free-flow travel time of a private or autonomous vehicle
on edge $(v,v')$. For a bus, it is computed with the lower bus free-flow
speed. Queuing delays at bus stops and bus dwell times are accounted for separately from these edge travel times.

\begin{figure}[!t]
    \centering
    \includegraphics[width=0.83\columnwidth]
    {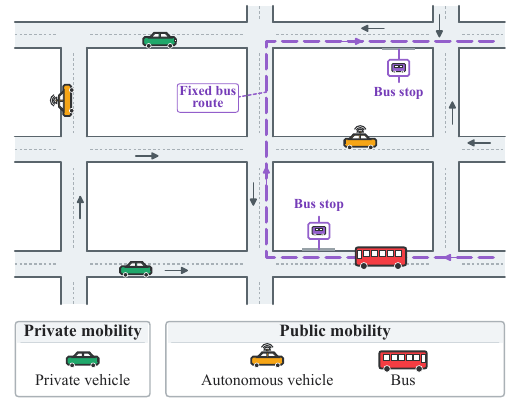}
    \vspace{-5pt}
    \caption{Illustration of the mixed traffic network with private vehicles, autonomous vehicles, and fixed-route buses.}
    \label{fig:mode_route_network}
\end{figure}

As shown in Fig.~\ref{fig:mode_route_network}, in this paper, we consider a mixed road traffic network consisting of \emph{private vehicles}, \emph{autonomous
vehicles}, and \emph{buses}. The corresponding travel
modes available to travelers are denoted as $\mu\in\{\mathrm{p},\mathrm{a},\mathrm{b}\}$, respectively. At any time $t$, let $N(t)\!\in\!\mathbb{N}$ denote the number of vehicles traveling in the network, with $\mathbb{N}$ being the set of nonnegative integers. Accordingly, the set of vehicles actively traveling in the road network is defined as $\mathcal{N}(t)\!:=\!\{1,2,\ldots,N(t)\}$. For each vehicle $i\!\in\!\mathcal{N}(t)$, let $r_i(t)$ denote its route, consisting of $n_i\!\geq\!2$ ordered nodes, where $n_i$ is the number of nodes on the route. The $k$-th node on the route is denoted by $v_{i,k}\!\in\!\mathcal{V}$, for $k\!=\!1,\dots,n_i$. The route is represented by the ordered sequence of directed edges
\begin{equation}
r_i(t)=\big\langle(v_{i,1},v_{i,2}),
(v_{i,2},v_{i,3}),\dots,(v_{i,n_i-1},v_{i,n_i})
\big\rangle,
\end{equation}
where $(v_{i,k},v_{i,k+1})\!\in\!\mathcal{E}$ for all $k\!=\!1,\dots,n_i\!-\!1$. 

In practice, private and autonomous vehicles typically have flexible routes that can be dynamically adjusted in response to real-time traffic conditions. We designate the intermediate road junctions along a route as decision-making points (DMPs), where vehicles can update their remaining routes based on the latest traffic information. Bus stops, trip origins, and destinations are not included in the set of DMPs. In contrast, buses follow preplanned routes that remain unchanged throughout their trips, serving bus stops in a prescribed order. To represent bus routes, let $\mathcal{L}$ denote the set of bus lines, and $r_\ell$ the ordered sequence of directed edges defining the fixed route of bus line $\ell\!\in\!\mathcal{L}$. 

\subsection{Vehicle Dynamics and Travel Time}
Next, we describe the dynamics of vehicles along their routes. For each
vehicle $i\!\in\!\mathcal{N}(t)$, let $a_{i,k}$ denote its arrival time at the
$k$-th node $v_{i,k}$ of its route, with $k\!=\!1,\ldots,n_i$, and let
$\delta_{i,k}\geq 0$ denote the additional time spent at node $v_{i,k}$ before
entering the next edge. If $v_{i,k}$ is a bus stop, $\delta_{i,k}$ includes
the queuing delay at the stop for all vehicle types and, if vehicle $i$ is a
bus, the dwell time for passenger boarding and alighting. These stop-related
delays are modeled separately from edge travel times~\cite{li2023transit}.
At all other nodes, $\delta_{i,k}\!=\!0$.

The dynamics of vehicle $i$ are described as
\begin{equation}
a_{i,k+1}=a_{i,k}+\delta_{i,k}+\tau(v_{i,k},v_{i,k+1}), \; k=1,\dots,n_i-1,
\label{eq:vehicle_arrival}
\end{equation}
where $\tau(v_{i,k},v_{i,k+1})$ denotes the travel time along the edge $(v_{i,k},v_{i,k+1})$, which is evaluated based on the traffic conditions when vehicle $i$ enters the edge.

We model the actual travel time along each edge using the widely adopted Bureau of Public Roads (BPR) function~\cite{wollenstein2022,gore2023modifiedbpr}. The BPR model captures
the impact of congestion by describing how travel time increases as traffic
demand approaches or exceeds road capacity. To account for the three vehicle
types sharing the road network, the actual travel time on edge $(v_{i,k},v_{i,k+1})$ is given by
\begin{equation}
\thinmuskip=2mu \medmuskip=2mu \thickmuskip=3mu
\tau(v_{i,k},v_{i,k+1})=\tau^{0}(v_{i,k},v_{i,k+1})
\Bigl[1+\alpha\Bigl(\tfrac{f_{i,k}+\eta_{\mathrm b}f_{i,k}^{\mathrm b}}{c_{i,k}}\Bigr)^{\!\beta}\Bigr],
\label{eq:mixed_bpr}
\end{equation}
where $f_{i,k}\geq0$ denotes the combined flow of private and autonomous
vehicles, and $f_{i,k}^{\mathrm b}\geq0$ denotes the bus flow on edge
$(v_{i,k},v_{i,k+1})$. Both flows are measured over the counting interval
preceding vehicle $i$'s entry into edge $(v_{i,k},v_{i,k+1})$. The factor $\eta_{\mathrm b}$ is the passenger car equivalent (PCE) of a bus,
defined as the number of passenger cars that produce the same effect on traffic
operations as one bus. Each private or autonomous vehicle counts as one PCE,
and we set \mbox{$\eta_{\mathrm b}\!=\!2$} following the heavy-vehicle equivalent
recommended in the Highway Capacity Manual~\cite{hcm2000}. Accordingly, the
road capacity $c_{i,k}>0$ of edge $(v_{i,k},v_{i,k+1})$ is measured in PCE per
hour. The positive parameters $\alpha$ and $\beta$ characterize the sensitivity
of travel time to traffic flow. In this paper, we employ the typical values
$\alpha\!=\!0.15$ and $\beta\!=\!4$.

Let $\mathcal{I}$ denote the set of all travelers considered. For traveler
$j\in\mathcal{I}$, let $T_j^\mu$ denote the total travel time under mode
\mbox{$\mu\in\{\mathrm p,\mathrm a,\mathrm b\}$}. If traveler $j$ travels in a
private or autonomous vehicle $i$, the total travel time is taken as the
travel time of that vehicle from its origin $v_{i,1}$ to its destination
$v_{i,n_i}$, i.e.,
\begin{equation}
T_j^\mu=a_{i,n_i}-a_{i,1},\qquad\mu\in\{\mathrm p,\mathrm a\}.
\label{eq:vehicle_total_travel_time}
\end{equation}
A bus traveler additionally walks to and from bus stops and waits for the
bus at the boarding stop. The total bus travel time is thus given by
\begin{equation}
T_j^{\mathrm b}=T_j^{\mathrm{walk}}+T_j^{\mathrm{wait}}+T_j^{\mathrm{ride}},
\label{eq:bus_total_travel_time}
\end{equation}
where $T_j^{\mathrm{walk}}$ is the walking time from the origin to the
boarding stop and from the alighting stop to the destination, and
$T_j^{\mathrm{wait}}$ is the waiting time at the boarding stop. The on-board
time $T_j^{\mathrm{ride}}$ spans from boarding to alighting. It consists of
the edge travel times of the bus and the queuing delays and dwell times at
the stops it serves en route, as described by~\eqref{eq:vehicle_arrival}.
Next, we formulate the mode recommendation and routing problem by defining
feasible travel modes and plans under service constraints.

\subsection{Mode Recommendation and Routing Problem}
\label{sec:mode_routing_problem}

\looseness=-1
We consider mode recommendation before departure and dynamic routing
guidance during travel. The traveler set $\mathcal{I}$ is partitioned into
two disjoint subsets, $\mathcal{I}^{\mathrm{c}}$ and $\mathcal{I}^{\mathrm{nc}}$,
representing travelers with and without private vehicles, respectively.
Travelers in $\mathcal{I}^{\mathrm{c}}$ retain their private vehicles,
whereas travelers in $\mathcal{I}^{\mathrm{nc}}$ receive a recommendation for
either autonomous vehicle or bus service. For traveler $j\in\mathcal{I}$, let
$o_j,d_j\in\mathcal{V}$ denote the origin and destination nodes, and let $t_j$
denote the time at which the traveler is ready to depart. The candidate mode
set of traveler $j$ is
\begin{equation}
\mathcal{M}_j=
\begin{cases}
\{\mathrm{p}\}, & j\in\mathcal{I}^{\mathrm{c}},\\[2pt]
\{\mathrm{a},\mathrm{b}\}, & j\in\mathcal{I}^{\mathrm{nc}}.
\end{cases}
\label{eq:feasible_modes}
\end{equation}
For private or autonomous vehicle travel, a travel plan $\rho$ specifies a
road route from $o_j$ to $d_j$. For bus travel, $\rho$ specifies a scheduled
bus trip, its boarding and alighting stops, and the walking paths to and from
these stops~\cite{salazar2020intermodal}. Let $\mathcal{R}_j^\mu(t)$ denote
the set of plans of mode $\mu$ available to traveler $j$ at time $t$.
Denoting a mode--plan pair by $u=(\mu,\rho)$, the candidate set of traveler
$j$ at time $t$ is
\begin{equation}
\Omega_j(t)=\left\{(\mu,\rho):\mu\in\mathcal{M}_j,\quad
\rho\in\mathcal{R}_j^\mu(t)\right\}.
\label{eq:feasible_mode_route_plans}
\end{equation}

A candidate pair in $\Omega_j(t)$ can be recommended only if it satisfies
the service constraints. Each autonomous vehicle serves one traveler for a
single trip without reuse, and the total number of autonomous vehicles
dispatched cannot exceed a prescribed limit. A bus plan must keep the total
walking distance within a prescribed limit, allow the traveler to reach the
boarding stop before the bus finishes serving it, and provide a seat on every
segment between the boarding and alighting stops. In addition, each
recommendation must preserve sufficient service capacity for travelers who
have known demand but have not yet received recommendations.

During travel, the recommended modes remain fixed. Private and autonomous vehicles start on prescribed initial routes and, update their remaining routes at each DMP, to minimize the predicted remaining travel time. Buses follow their fixed routes, and bus
travelers remain on their selected trips without transfers. This paper aims
to recommend a feasible mode--plan pair $u_j^*\in\Omega_j(t_j)$ for each
traveler $j\in\mathcal{I}^{\mathrm{nc}}$. Combined with dynamic routing
guidance for private and autonomous vehicles, these recommendations are
intended to improve the average trip quality of all travelers in
$\mathcal{I}$. Next, we introduce the DTX to evaluate trip quality
consistently across travel modes.

\section{Dynamic Trip Index}
\label{sec:dynamic_trip_index}

In this section, we formulate the DTX, which incorporates mode-specific
travel time, monetary cost, and convenience. The index provides a consistent
measure of trip quality across private vehicles, autonomous vehicles, and
buses, and serves as the basis for mode recommendation.

Consider traveler $j\in\mathcal{I}$ with origin--destination pair
$(o_j,d_j)$ and a mode--plan pair $u=(\mu,\rho)$ as defined
above. Building on the trip-index
formulation in~\cite{bai2025routing}, we define the DTX of traveler $j$
under $u$ as
\begin{equation}
\mathrm{DTX}_{j}(u)
=
\xi_{1}\frac{T_{\min}^{j}}{T_{j}(u)}
+
\xi_{2}\frac{C_{\min}^{j}}{C_{j}(u)}
+
\xi_{3}\frac{Q_{\min}^{j}}{Q_{\mu}}.
\label{eq:dtx}
\end{equation}
The three terms evaluate the efficiency, cost, and convenience of the trip,
respectively. The weights $\xi_1,\xi_2,\xi_3\in\mathbb{R}_+$ are positive
constants that satisfy the convex combination condition
$\xi_1+\xi_2+\xi_3=1$. In the efficiency term, $T_j(u)>0$ denotes the total
travel time of traveler $j$ under $u$, which equals $T_j^\mu$ given
by~\eqref{eq:vehicle_total_travel_time} for $\mu\in\{\mathrm p,\mathrm a\}$
and by~\eqref{eq:bus_total_travel_time} for $\mu=\mathrm b$, and
$T_{\min}^{j}$ denotes the minimum total travel time of the trip. In the cost
term, $C_j(u)>0$ denotes the monetary cost under $u$, and $C_{\min}^{j}$
denotes the minimum monetary cost of the trip. In the convenience term,
$Q_\mu>0$ denotes the inconvenience of mode $\mu$, and $Q_{\min}^{j}$ denotes
the minimum inconvenience of the trip. A larger DTX indicates higher trip
quality. The three minima are determined over all travel modes and their
plans under ideal traffic conditions, i.e., without congestion or queuing
delays, and are computed as
\begin{subequations}
\label{eq:dtx_references}
\begin{align}
T_{\min}^{j}
&=\min_{r\in\mathcal{R}_{o_j,d_j}}
\sum_{(v,v')\in r}\tau^{0}(v,v'),
\label{eq:dtx_time_reference}\\
C_{\min}^{j}
&=\min\left\{\epsilon_{\min}T_{\min}^{j},E_{\mathrm b}\right\},
\label{eq:dtx_cost_reference}\\
Q_{\min}^{j}
&=\frac{T_{w,\min}^{j}}{T_{d,\max}^{j}},
\label{eq:dtx_inconvenience_reference}
\end{align}
\end{subequations}
where $\mathcal{R}_{o_j,d_j}$ denotes the set of road routes from $o_j$ to
$d_j$ in $\mathcal{G}$, and $\tau^{0}(v,v')$ is the free-flow travel time on
edge $(v,v')$. Thus,
\eqref{eq:dtx_time_reference} gives the shortest free-flow driving time of
the trip. In~\eqref{eq:dtx_cost_reference}, $\epsilon_{\mathrm p}$ and
$\epsilon_{\mathrm a}$ denote the costs per unit travel time of private and
autonomous vehicles, respectively,
$\epsilon_{\min}=\min\{\epsilon_{\mathrm p},\epsilon_{\mathrm a}\}$, and
$E_{\mathrm b}$ denotes the fixed fare of a single bus trip. Hence,
$C_{\min}^{j}$ is the smaller of the minimum vehicle cost over
$T_{\min}^{j}$ and the bus fare. In~\eqref{eq:dtx_inconvenience_reference},
the minimum average preparation time and the maximum daily service duration
are
\begin{subequations}
\label{eq:convenience_reference_components}
\begin{align}
T_{w,\min}^{j}
&=\min\left\{T_{w,\mathrm p},T_{w,\mathrm a},T_{w,\mathrm b}\right\},
\label{eq:minimum_additional_wait}\\
T_{d,\max}^{j}
&=\max\left\{T_{d,\mathrm p},T_{d,\mathrm a},T_{d,\mathrm b}\right\}.
\label{eq:maximum_departure_window}
\end{align}
\end{subequations}
{Here, $T_{w,\mu}$ denotes the average preparation time that a traveler needs before a trip with mode $\mu$ can start, and $T_{d,\mu}$ denotes the daily service duration of mode $\mu$, which is the number of hours per day during which such a trip can be started. A private vehicle is available at any time of day, and $T_{w,\mathrm p}$ accounts for the time required to retrieve it from parking. An autonomous vehicle is available during the operating hours, and $T_{w,\mathrm a}$ accounts for the time from the trip request until the vehicle is dispatched. The buses considered in this paper serve peak hour lines, which operate only for a few hours during the morning and evening peaks and offer a small number of scheduled trips. A bus traveler therefore has to adjust the planned departure to the timetable in advance, and $T_{w,\mathrm b}$ accounts for this adjustment. These preparation times occur before the trip starts and are not included in the total travel times in~\eqref{eq:vehicle_total_travel_time} and~\eqref{eq:bus_total_travel_time}. In particular, $T_{w,\mathrm b}$ differs from the waiting time $T_j^{\mathrm{wait}}$ at the boarding stop. A mode that requires a longer preparation time or is available for fewer hours per day is less convenient for travelers.} The monetary cost and
inconvenience in~\eqref{eq:dtx} are computed as
\begin{align}
C_j(u)
&=\begin{cases}
\epsilon_\mu T_j(u), & \mu\in\{\mathrm p,\mathrm a\},\\[2pt]
E_{\mathrm b}, & \mu=\mathrm b,
\end{cases}
\label{eq:dtx_cost}\\
Q_\mu
&=\frac{T_{w,\mu}}{T_{d,\mu}},\qquad
\mu\in\{\mathrm p,\mathrm a,\mathrm b\},
\label{eq:dtx_inconvenience}
\end{align}
where $\epsilon_\mu$, $T_{w,\mu}$, and $T_{d,\mu}$ take the values
corresponding to the selected mode $\mu$. The parameter values used in this
study are specified with the simulation setup. In the next section, we present a method that integrates
DTX-based mode recommendation with dynamic routing guidance.

\section{Mode Recommendation and Dynamic Routing Guidance}
\label{sec:mode_recommendation}

This section presents the integrated mode recommendation and dynamic routing
guidance method for the problem formulated
above. Before departure, the method
recommends to each traveler $j\in\mathcal{I}^{\mathrm{nc}}$ the mode--plan
pair that maximizes the DTX predicted from current traffic and service
information, subject to the service constraints. During travel, it updates
the remaining routes of private and autonomous vehicles at each DMP to
minimize the predicted remaining travel time.

\subsection{DTX-Based Mode Recommendation}
\label{sec:dtx_mode_recommendation}
\setcounter{topnumber}{1}

Travelers in $\mathcal{I}^{\mathrm{nc}}$ receive recommendations in the order
of their ready times $t_j$. Since the realized DTX of a candidate pair is
unknown before departure, traveler $j\!\in\!\mathcal{I}^{\mathrm{nc}}$ is
recommended
\begin{equation}
u_j^*\in\underset{u\in\mathcal{F}_j}{\arg\max}\;\widehat{\mathrm{DTX}}_j(u),
\label{eq:dtx_recommendation}
\end{equation}
where $\widehat{\mathrm{DTX}}_j(u)$ is the predicted DTX of pair $u$, and
$\mathcal{F}_j\subseteq\Omega_j(t_j)$ is the set of candidate pairs that
satisfy the service constraints at time $t_j$. {Problem~\eqref{eq:dtx_recommendation} thus maximizes the predicted DTX of each traveler in turn, given earlier commitments, rather than the mean DTX over $\mathcal{I}$, whose improvement is evaluated through simulation.}

\subsubsection{Trip Attribute Prediction}
\label{sec:trip_attribute_prediction}
Predictions use only the information available at time $t_j$. Road traffic
estimates are refreshed at the window boundaries $t_g=t_0+g\Delta_{\mathrm w}$,
$g\in\mathbb{N}$, where $t_0$ is the start of the recommendation period and
$\Delta_{\mathrm w}>0$ is the window length. These estimates remain fixed
within each window $[t_g,t_{g+1})$. Let $\mathcal{V}^{\mathrm b}\subseteq
\mathcal{V}$ denote the set of bus-stop nodes. At $t_g$, the predicted travel
time $\widehat\tau_g(v,v')$ of each edge $(v,v')\in\mathcal{E}$ is obtained
from~\eqref{eq:mixed_bpr} using the flows counted in the interval preceding
$t_g$. The predicted queuing delay $\widehat\delta_g(v)$ at each node
$v\in\mathcal{V}^{\mathrm b}$ is estimated from its current queue, and
$\widehat\delta_g(v)=0$ for $v\in\mathcal{V}\setminus\mathcal{V}^{\mathrm b}$.
Bus positions and service states are updated continuously and are used as
observed at $t_j$.

For an autonomous vehicle plan $u=(\mathrm a,\rho)$ with
$\rho\in\mathcal{R}_j^{\mathrm a}(t_j)$ and $t_j\in[t_g,t_{g+1})$, the
predicted total travel time and monetary cost follow
from~\eqref{eq:vehicle_arrival}, \eqref{eq:vehicle_total_travel_time},
and~\eqref{eq:dtx_cost} as
\begin{align}
\widehat T_j(u)&=\sum_{(v,v')\in\rho}
\big[\widehat\delta_g(v)+\widehat\tau_g(v,v')\big],
\label{eq:av_predicted_time}\\
\widehat C_j(u)&=\epsilon_{\mathrm a}\widehat T_j(u).
\label{eq:av_predicted_cost}
\end{align}
Both the efficiency and cost terms in~\eqref{eq:dtx} decrease as
$\widehat T_j(u)$ increases, whereas the convenience term is fixed.
{Since trip origins are not DMPs, the vehicle follows its prescribed
initial route until the first DMP on that route. Hence, only one autonomous
vehicle plan $\rho_j^{\mathrm a}$ needs to be evaluated, which follows the
prescribed initial route up to this DMP and then the remaining route with the
shortest predicted travel time.}

For each bus line $\ell\in\mathcal{L}$, let $S_\ell$ be the number of bus
stops on its route $r_\ell$, indexed by $s=1,\dots,S_\ell$ in the order of
service, and let $\Theta_\ell$ denote the set of scheduled trips of line
$\ell$, each of which serves these stops in order. For a bus plan
$u=(\mathrm b,\rho)$ with $\rho\in\mathcal{R}_j^{\mathrm b}(t_j)$, $\rho$
specifies a trip $\theta\in\Theta_\ell$ of some line $\ell\in\mathcal{L}$, the
indices $s^{\mathrm{begin}},s^{\mathrm{end}}\in\{1,\dots,S_\ell\}$ of the boarding
and alighting stops with $s^{\mathrm{begin}}<s^{\mathrm{end}}$, and the walking
paths to the boarding stop and from the alighting stop. The walking times
along these two paths are $w_{j,1}(\rho)\geq0$ and $w_{j,2}(\rho)\geq0$,
respectively, and the total walking distance is $d_j(\rho)\geq0$. Starting from the observed bus states at $t_j$, the service
start times of trip $\theta$ at its subsequent stops are forecast
sequentially using the road estimates of window $g$. Let
$\widehat t_\theta(s)$ denote the predicted service start time of trip
$\theta$ at stop $s$. The traveler boards at
$\widehat t_j^{\mathrm{board}}(\rho)=\max\{t_j+w_{j,1}(\rho),
\widehat t_\theta(s^{\mathrm{begin}})\}$ and alights at
$\widehat t_\theta(s^{\mathrm{end}})$. Accordingly, the three components
in~\eqref{eq:bus_total_travel_time} are predicted as
\begin{subequations}
\label{eq:bus_prediction}
\begin{align}
\widehat T_j^{\mathrm{walk}}(u)&=w_{j,1}(\rho)+w_{j,2}(\rho),
\label{eq:bus_predicted_walk}\\
\widehat T_j^{\mathrm{wait}}(u)&=\widehat t_j^{\mathrm{board}}(\rho)-t_j
-w_{j,1}(\rho),
\label{eq:bus_predicted_wait}\\
\widehat T_j^{\mathrm{ride}}(u)&=\widehat t_\theta(s^{\mathrm{end}})
-\widehat t_j^{\mathrm{board}}(\rho),
\label{eq:bus_predicted_ride}
\end{align}
\end{subequations}
and the predicted total travel time $\widehat T_j(u)$ is their sum, with
$\widehat C_j(u)=E_{\mathrm b}$. For both modes, substituting
$\widehat T_j(u)$ and $\widehat C_j(u)$ into~\eqref{eq:dtx} gives
$\widehat{\mathrm{DTX}}_j(u)$, where $T_{\min}^{j}$, $C_{\min}^{j}$, and
$Q_{\min}^{j}$ remain those given by~\eqref{eq:dtx_references} under ideal
traffic conditions.

\begin{algorithm}[t]
\caption{DTX-Based Mode Recommendation}
\label{alg:windowed_dtx_recommendation}
\footnotesize
\KwIn{$\mathcal{G}$, $\mathcal{I}^{\mathrm{nc}}$,
$\{o_j,d_j,t_j\}_{j\in\mathcal{I}^{\mathrm{nc}}}$, $\mathcal{H}$, $A^{\max}$,
$\{B_\theta\}$}
\KwOut{$\{u_j^*\}_{j\in\mathcal{I}^{\mathrm{nc}}}$}
Initialization: $A^{\mathrm{used}}\gets0$, $b_{\theta,s}\gets0$ for all
$\theta,s$, {and $\rho_j^{\mathrm a}\gets$ prescribed initial route}\;
\For{$j\in\mathcal{I}^{\mathrm{nc}}$ in ascending order of $t_j$}{
  obtain $\widehat\tau_g,\widehat\delta_g$ with $t_j\in[t_g,t_{g+1})$\;
  $\mathcal{F}_j\gets\emptyset$\;
  \If{\eqref{eq:av_protection} holds}{
    {\SetAlgoHangIndent{0pt}%
    {update $\rho_j^{\mathrm a}$ to follow the route with the shortest
    predicted travel time}\;}
    $\mathcal{F}_j\gets\mathcal{F}_j\cup\{(\mathrm a,{\rho_j^{\mathrm a}})\}$\;
  }
  \For{$\rho\in\mathcal{R}_j^{\mathrm b}(t_j)$}{
    \If{$d_j(\rho)\leq d^{\max}$ \textbf{and} \eqref{eq:bus_catchability},
    \eqref{eq:bus_protection} hold}{
      $\mathcal{F}_j\gets\mathcal{F}_j\cup\{(\mathrm b,\rho)\}$\;
    }
  }
  \eIf{$\mathcal{F}_j=\emptyset$}{
    report recommendation failure for $j$\;
  }{
    compute $\widehat{\mathrm{DTX}}_j(u)$, $u\in\mathcal{F}_j$,
    by \eqref{eq:dtx}, \eqref{eq:av_predicted_time}--\eqref{eq:bus_prediction}\;
    select $u_j^*$ by \eqref{eq:dtx_recommendation} with tie-breaking\;
    remove the record of $j$ from $\mathcal{H}$\;
    \lIf{$u_j^*=(\mathrm a, {\rho_j^{\mathrm a}})$}{$A^{\mathrm{used}}\gets
    A^{\mathrm{used}}+1$}
    \lElse{$b_{\theta,s}\gets b_{\theta,s}+1$ for
    $s^{\mathrm{begin}}\leq s<s^{\mathrm{end}}$}
  }
}
\Return $\{u_j^*\}_{j\in\mathcal{I}^{\mathrm{nc}}}$
\end{algorithm}

\subsubsection{Service Constraints}
\label{sec:service_constraints}
A candidate pair is included in $\mathcal{F}_j$ only if it satisfies the
service constraints. A bus plan must satisfy $d_j(\rho)\leq d^{\max}$, where
$d^{\max}>0$ is the walking-distance limit. In addition, the traveler must
reach the boarding stop before the bus finishes serving it, i.e.,
\begin{equation}
t_j+w_{j,1}(\rho)+\gamma\leq
\underline{t}_\theta^{\mathrm{leave}}(s^{\mathrm{begin}}),
\label{eq:bus_catchability}
\end{equation}
where $\gamma\geq0$ is a safety margin, and
$\underline{t}_\theta^{\mathrm{leave}}(s^{\mathrm{begin}})$ is a lower bound,
indicated by the underline, on the time at which trip $\theta$ leaves stop
$s^{\mathrm{begin}}$. This
bound is computed from the observed bus state using free-flow travel times
and fixed dwell times.

The remaining constraints concern limited service capacity. If each traveler
simply took the best available resource, early travelers could exhaust the
autonomous vehicles or bus seats needed by later travelers with known demand.
To ensure that no traveler fails to be served because of insufficient
capacity, we maintain a protection set $\mathcal{H}$ that contains one record
for each traveler in $\mathcal{I}^{\mathrm{nc}}$ who has not yet received a
recommendation. Each record reserves either one autonomous vehicle or one bus
seat between specified boarding and alighting stops. The records are initialized
from the travelers' initial service assignments in the known demand, which
satisfy the capacity limits. They reserve capacity without fixing the
recommended modes.

Let $A^{\max}\in\mathbb{N}$ be the maximum number of autonomous vehicles that
can be dispatched, and let $A^{\mathrm{used}}\in\mathbb{N}$ be the number
already used by previous recommendations, which is initially zero. Let
$P_{-j}^{\mathrm a}\in\mathbb{N}$ be the number of autonomous vehicle records
in $\mathcal{H}$ other than that of traveler $j$. An autonomous vehicle plan
is admissible only if
\begin{equation}
A^{\mathrm{used}}+P_{-j}^{\mathrm a}+1\leq A^{\max}.
\label{eq:av_protection}
\end{equation}
\looseness=-1
Since each autonomous vehicle serves a single trip without reuse, completed
trips do not reduce $A^{\mathrm{used}}$. For trip $\theta\in\Theta_\ell$, let
$B_\theta\in\mathbb{N}$ be its seat capacity, and let segment
$s\in\{1,\dots,S_\ell-1\}$ denote the part of the route between stops $s$ and
$s+1$. Let $b_{\theta,s}\in\mathbb{N}$ be the number of seats committed on
segment $s$, which is initially zero, and let
$P_{\theta,s,-j}^{\mathrm b}\in\mathbb{N}$ be the number of records in
$\mathcal{H}$ other than that of traveler $j$ that reserve a seat on this
segment. A bus plan is admissible only if
\begin{equation}
b_{\theta,s}+P_{\theta,s,-j}^{\mathrm b}+1\leq B_\theta,\qquad
s^{\mathrm{begin}}\leq s<s^{\mathrm{end}}.
\label{eq:bus_protection}
\end{equation}
The set $\mathcal{F}_j$ thus consists of the candidates in $\Omega_j(t_j)$
that satisfy~\eqref{eq:av_protection} for autonomous vehicle plans, or the
walking-distance limit, \eqref{eq:bus_catchability},
and~\eqref{eq:bus_protection} for bus plans.

\subsubsection{Recommendation and Commitment}
\label{sec:recommendation_commitment}
The recommended pair $u_j^*$ is obtained by
solving~\eqref{eq:dtx_recommendation} over $\mathcal{F}_j$. If several pairs
attain the maximum, ties are broken in favor of the larger slack
in~\eqref{eq:av_protection} or~\eqref{eq:bus_protection}, i.e., the
difference between the right- and left-hand sides, taken as the minimum over
the ridden segments for bus plans. Remaining ties favor bus plans over
autonomous vehicle plans and are finally resolved by a fixed order of plan
attributes. If $\mathcal{F}_j$ is empty, the recommendation fails, and all
resources remain unchanged. Otherwise, the record of traveler $j$ is removed
from $\mathcal{H}$, and the selected resource is committed. An autonomous
vehicle selection increases $A^{\mathrm{used}}$ by one, whereas a bus
selection increases $b_{\theta,s}$ by one on every ridden segment
$s^{\mathrm{begin}}\leq s<s^{\mathrm{end}}$.
Algorithm~\ref{alg:windowed_dtx_recommendation} summarizes the procedure.

\subsection{Dynamic Routing Guidance}
\label{sec:dynamic_routing}
During travel, the modes of all travelers remain fixed, and the routes of
private and autonomous vehicles are updated at DMPs. Consider a private or
autonomous vehicle $i\in\mathcal{N}(t)$ that arrives at a DMP $v_{i,k}$ at
time $t=a_{i,k}$. Let $\mathcal{R}_{i,k}(t)$ denote the set of road routes
from $v_{i,k}$ to the destination $v_{i,n_i}$ that are accessible to vehicle
$i$ and comply with the permitted turns from the incoming edge
$(v_{i,k-1},v_{i,k})$. Using the DMP-based route update in~\cite{bai2025routing}
with a travel-time objective, the remaining route of vehicle $i$ is
determined by solving
\begin{equation}
\min_{r\in\mathcal{R}_{i,k}(t)}\;\widehat T_{i,k}(r)=\sum_{(v,v')\in r}
\big[\widehat\delta_t(v)+\widehat\tau_t(v,v')\big],
\label{eq:routing_objective}
\end{equation}
where $\widehat T_{i,k}(r)$ is the predicted remaining travel time along
route $r$. This expression follows from applying~\eqref{eq:vehicle_arrival}
along $r$ from $a_{i,k}$, with the queuing delays and edge travel times
replaced by their predictions at time $t$.

The predictions use only the traffic information available at time $t$. The
predicted travel time $\widehat\tau_t(v,v')$ of each edge
$(v,v')\in\mathcal{E}$ is obtained from~\eqref{eq:mixed_bpr} using the flows
counted in the most recent counting interval before $t$, with zero flows
initially. Unlike the predictions for mode recommendation, which are
refreshed once per window, these estimates are updated at every counting
interval. The predicted queuing delay $\widehat\delta_t(v)$ at each node
$v\in\mathcal{V}^{\mathrm b}$ is the delay that vehicle $i$ would experience
if it joined the end of the current queue at stop $v$, accounting for the bus
currently being served there, and $\widehat\delta_t(v)=0$ for
$v\in\mathcal{V}\setminus\mathcal{V}^{\mathrm b}$. These predictions remain
fixed while~\eqref{eq:routing_objective} is solved.

Since all predicted delays and edge travel times are nonnegative and fixed,
\eqref{eq:routing_objective} is a shortest-path problem and is solved by
Dijkstra's algorithm while respecting the permitted turns. Let
$r_{i,k}^{\mathrm{cur}}$ denote the untraveled part of the current route of
vehicle $i$. If $r_{i,k}^{\mathrm{cur}}$ attains the minimum
in~\eqref{eq:routing_objective}, it is retained. Otherwise, it is replaced by
a minimizer of~\eqref{eq:routing_objective}. The route update changes only the untraveled part, so the
traversed route and the destination of vehicle $i$ remain unchanged. The
same procedure is repeated at each subsequent DMP.

\begin{table}[t]
\caption{Parameter Settings of Each Travel Mode}
\label{tab:dtx_parameters}
\centering
\footnotesize
\setlength{\tabcolsep}{2pt}
\renewcommand{\arraystretch}{1.15}
\begin{tabular*}{\columnwidth}{@{\extracolsep{\fill}}lccccccc@{}}
\toprule
Mode $\mu$ & $\xi_1$ & $\xi_2$ & $\xi_3$
& $\epsilon_\mu$ & $E_{\mathrm{b}}$ & $T_{w,\mu}$ & $T_{d,\mu}$\\
& & & & [\$/min] & [\$/trip] & [min] & [h/day]\\
\midrule
Private vehicle $\mathrm{p}$ & 0.4 & 0.4 & 0.2 & 0.1791 & -- & 2 & 24\\
Autonomous vehicle $\mathrm{a}$ & 0.4 & 0.4 & 0.2 & 0.0985 & -- & 15 & 18\\
Bus $\mathrm{b}$ & 0.4 & 0.4 & 0.2 & -- & 0.2985 & 25 & 3\\
\bottomrule
\end{tabular*}
\end{table}

\section{Simulation Studies}
\label{sec:simulation_studies}
\setlength{\textfloatsep}{8pt plus 2pt minus 2pt}
\setlength{\floatsep}{8pt plus 2pt minus 2pt}
\setcounter{topnumber}{2}
\setcounter{totalnumber}{2}
\renewcommand{\topfraction}{0.90}
\renewcommand{\textfraction}{0.08}
\renewcommand{\floatpagefraction}{0.85}

\subsection{Simulation Setup}
\label{sec:simulation_setup}

We consider an urban road network in Shanghai, China,
with $57$ road junctions, $24$ bus stops, and $130$ directed
roads. Splitting the roads at bus stops results in $154$
directed edges. The simulation involves $|\mathcal{I}|=1{,}500$
travelers with randomly assigned departure times between
08:00 and 09:00, corresponding to $[0,3600)$~s. Among them,
$625$ own private vehicles, while the remaining $875$ are
initially assigned to $625$ autonomous vehicle trips and
$250$ bus trips. Private and autonomous vehicles travel at
a free-flow speed of $30$~km/h, buses at $20$~km/h, and
travelers walk at $5$~km/h. The free-flow travel time of
each edge is calculated from its length and the corresponding
vehicle speed. The road capacity is set to $300$~PCE/h, and
edge flows are counted every $60$~s. Two bidirectional bus
lines operate, each with five scheduled trips in each direction
at a headway of $15$~min. At most $A^{\max}=625$
autonomous vehicles can be dispatched, with each vehicle
serving a single trip without reuse. The simulation terminates
once all trips are completed. The weights in
Table~\ref{tab:dtx_parameters} satisfy
$\xi_1+\xi_2+\xi_3=1$.

We compare the proposed framework with two baselines under
the same travel demand and initial vehicle routes:
\begin{itemize}
    \item \emph{Fixed plans}: All travelers retain their initial
    travel modes, and all vehicles follow their prescribed
    initial routes throughout their trips.

    \item \emph{Routing only}: All travelers retain their initial
    travel modes, while private and autonomous vehicles update
    their routes at DMPs using dynamic routing guidance.

    \item \emph{Proposed method}: Travelers in
    $\mathcal{I}^{\mathrm{nc}}$ receive mode recommendations
    using Algorithm~\ref{alg:windowed_dtx_recommendation},
    while private and autonomous vehicles receive the same
    dynamic routing guidance as in Routing only.
\end{itemize}
In all three methods, bus routes and schedules remain fixed,
and each autonomous vehicle serves a single trip without reuse.
Travel modes remain unchanged once assigned before departure.

For each traveler $j\!\in\!\mathcal{I}$, the realized DTX,
denoted by $\mathrm{DTX}_j^{\mathrm{real}}$, is calculated
from~\eqref{eq:dtx} using the realized total travel time and
monetary cost, together with the assigned travel mode. The
reference values $T_{\min}^{j}$, $C_{\min}^{j}$, and
$Q_{\min}^{j}$ remain fixed as defined in
\eqref{eq:dtx_references}. {If a traveler cannot be
assigned a feasible mode--plan pair, the realized DTX is set
to zero, and the traveler is retained in $\mathcal{I}$ when
computing the mean. Thus, recommendation failures directly
reduce the mean realized DTX. No recommendation failures
occur in the simulation, and all $1{,}500$ travelers complete
their trips under each method.} The mean realized DTX is
\begin{equation}
\overline{\mathrm{DTX}}^{\mathrm{real}}
=\frac{1}{|\mathcal{I}|}\sum_{j\in\mathcal{I}}\mathrm{DTX}_j^{\mathrm{real}}.
\label{eq:mean_realized_dtx}
\end{equation}
As a supplementary measure of trip quality distribution,
we examine the dynamic trip equity (DTE)~\cite{bai2025routing},
defined as
\begin{equation}
\mathrm{DTE}=1-\frac{
\displaystyle\sum_{j\in\mathcal{I}}\sum_{j'\in\mathcal{I}}
\left|\mathrm{DTX}_j^{\mathrm{real}}-\mathrm{DTX}_{j'}^{\mathrm{real}}\right|
}{2|\mathcal{I}|^2\,\overline{\mathrm{DTX}}^{\mathrm{real}}},
\label{eq:posthoc_dte}
\end{equation}
where $\mathrm{DTE}=1$ indicates identical realized DTX
values across all travelers. For individual comparisons,
$\Delta\mathrm{DTX}_j$ denotes the difference between the
realized DTX of traveler $j$ under a given method and that
under Fixed plans.

\begin{figure}[t]
\centering
\includegraphics[width=0.78\columnwidth]{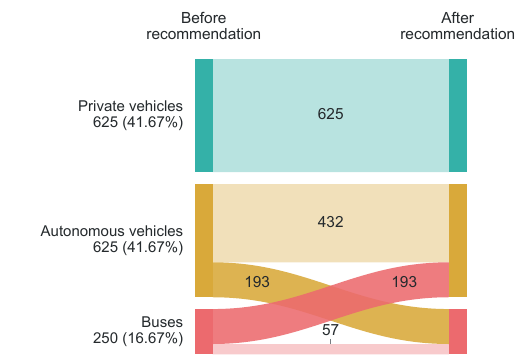}
\caption{Travel-mode shares before and after recommendation in the proposed method.
Bands show the numbers of travelers who kept or changed their modes.}
\label{fig:mode_transitions}
\end{figure}

\subsection{Results and Analysis}
\label{sec:simulation_results}

Table~\ref{tab:overall_results} shows that the proposed method achieves the highest mean realized DTX, with improvements of $19.68\%$ over Fixed plans and $13.01\%$ over Routing only. These results demonstrate the additional benefits of mode recommendation beyond route adaptation. Compared with Fixed plans, the proposed method reduces mean total travel time and monetary cost by $15.21\%$ and $37.68\%$, respectively. The corresponding reductions relative to Routing only are $8.71\%$ and $32.80\%$. Moreover, the DTE increases from $0.8011$ under Fixed plans to $0.8947$ under the proposed method, indicating a more even distribution of trip quality.

Fig.~\ref{fig:mode_transitions} illustrates the travel-mode shares before and after recommendation. Among the $625$ travelers initially assigned to autonomous vehicles, $432$ retain their modes, while $193$ switch to buses. Among the $250$ travelers initially assigned to buses, $57$ retain their modes, while $193$ switch to autonomous vehicles. Thus, a total of $386$ travelers, accounting for $25.73\%$ of all travelers, change modes. Since the number of travelers switching in each direction is equal, the aggregate shares of autonomous vehicles and buses remain unchanged.

\begin{figure}[t]
\centering
\includegraphics[width=0.75\columnwidth]{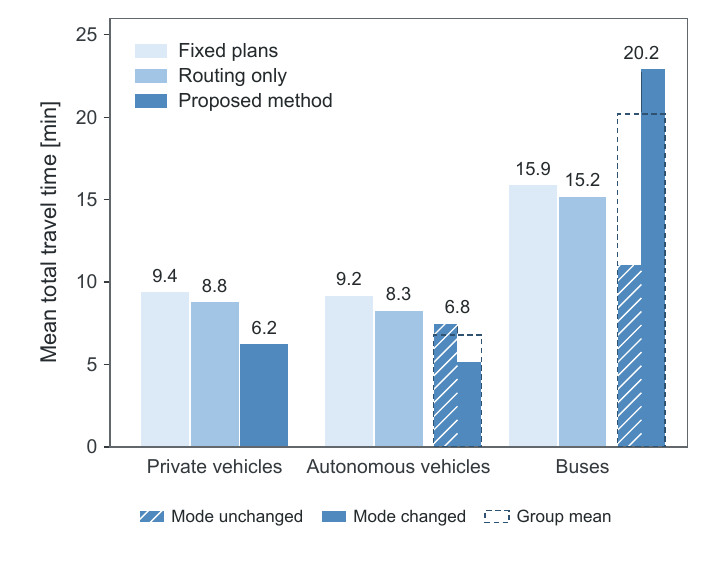}
\vspace{-10pt}
\caption{Mean travel time by travel mode after recommendation. For the proposed method, hatched and solid bars denote travelers whose modes are unchanged and changed by the recommendation, respectively, and dashed outlines denote the mean over all travelers using each mode.}
\label{Fig.3}
\end{figure}

\begin{figure}[t]
\centering
\includegraphics[width=0.75\columnwidth]{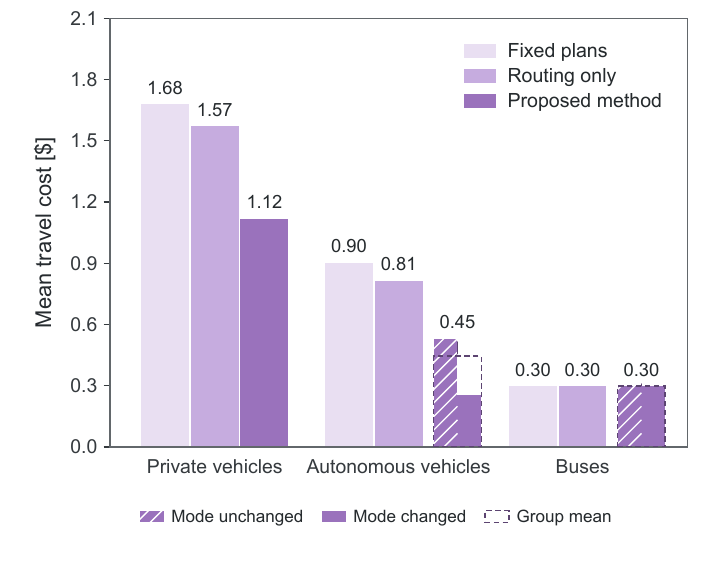}
\vspace{-10pt}
\caption{Mean monetary cost by travel mode after recommendation.}
\label{Fig.4}
\end{figure}


Fig.~\ref{Fig.3} compares mean total travel time across travel modes. The proposed method reduces the mean travel time of private vehicle users to $6.2$~min, compared with $9.4$~min under Fixed plans and $8.8$~min under Routing only, with a corresponding reduction in monetary cost, as shown in Fig.~\ref{Fig.4}. Since these travelers retain their modes and receive the same routing guidance as under Routing only, the improvement results from reduced congestion due to the redistribution of other travelers across modes. For bus users, the mean travel time increases from $15.9$ to $20.2$~min, primarily due to changes in traveler composition. Specifically, $193$ travelers switch from buses to autonomous vehicles, reducing their mean travel time from $16.2$ to $5.2$~min, while the $57$ travelers who remain on buses also benefit from a reduction from $14.8$ to $11.0$~min. In contrast, the $193$ travelers switching from autonomous vehicles to buses experience an increase in mean travel time from $11.6$ to $22.9$~min, but their mean monetary cost decreases from $\$1.14$ to $\$0.30$, resulting in an increase in mean realized DTX from $0.358$ to $0.498$.

\begin{figure}[t]
\centering
\setlength{\abovecaptionskip}{4pt}
\includegraphics[width=0.86\columnwidth,trim=0 3mm 0 0,clip]{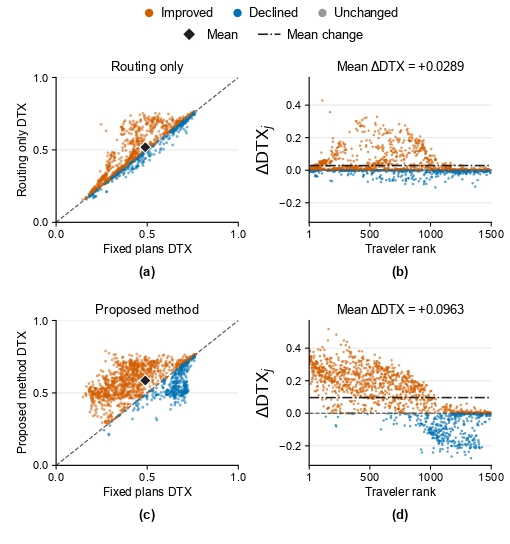}
\vspace{-5pt}
\caption{Individual realized DTX relative to Fixed plans. (a), (c): DTX under
Routing only and the proposed method versus that under Fixed plans. Points above the
diagonal indicate improvement, and diamonds denote mean DTX. (b), (d):
$\Delta\mathrm{DTX}_j$ sorted by DTX under Fixed plans. Dash-dotted lines
denote mean changes.}
\label{fig:paired_dtx}
\end{figure}

\begin{figure}[t]
\centering
\includegraphics[width=0.8\columnwidth]{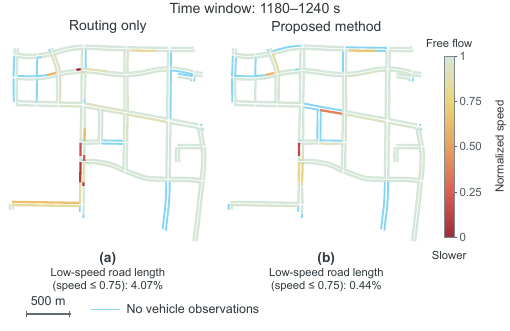}
\vspace{-10pt}
\caption{Normalized speed on each road during $[1180,1240)$~s under
(a) Routing only and (b) the proposed method. Red indicates slower traffic, and cyan
marks roads without vehicle observations.}
\label{fig:network_heatmap}
\end{figure}

Fig.~\ref{fig:paired_dtx} shows that the proposed method improves trip quality for most travelers. A total of $1{,}077$ travelers, or $71.80\%$, achieve a higher realized DTX than under Fixed plans, compared with $933$ under Routing only. The mean $\Delta\mathrm{DTX}_j$ reaches $0.0963$, more than three times that under Routing only, although $421$ travelers experience a decrease in DTX. Fig.~\ref{fig:network_heatmap} further illustrates the reduction in network congestion. The normalized road speed is defined as the ratio of free-flow to actual travel time, including queuing delays and bus dwell times, with $1$ indicating free flow. During $[1180,1240)$~s, roads with a normalized speed no greater than $0.75$ account for $4.07\%$ of the total road length under Routing only, compared with $0.44\%$ under the proposed method.

To examine the sensitivity of the proposed method to traveler preferences, we consider three alternative weight configurations: equal weights, a higher weight on convenience, and a higher weight on monetary cost. As shown in Table~\ref{tab:sensitivity}, the proposed method outperforms both baselines under all three configurations. The DTX gains are smaller when equal or greater importance is assigned to convenience, while the mean total travel time remains lower than under both baselines. In contrast, assigning a greater weight to monetary cost leads to more travelers being recommended bus services, resulting in larger DTX gains at the expense of longer mean total travel time. These results illustrate how the proposed method adapts mode recommendations to different preferences while consistently improving mean trip quality.

\suppressfloats[t]
\begin{table}[t]
\caption{Comparison with Baselines}
\label{tab:overall_results}
\centering
\small
\setlength{\tabcolsep}{3pt}
\renewcommand{\arraystretch}{1.15}
\begin{tabular*}{\columnwidth}{@{\extracolsep{\fill}}lcccc@{}}
\toprule
Method & DTX $\uparrow$ & Time [s] $\downarrow$ & Cost [\$] $\downarrow$ & DTE $\uparrow$\\
\midrule
Fixed plans & 0.4895 & 621.84 & 1.125 & 0.8011\\
Routing only & 0.5184 & 577.56 & 1.043 & 0.8140\\
\textbf{Proposed method} & \textbf{0.5859} & \textbf{527.25} & \textbf{0.701} & \textbf{0.8947}\\
\bottomrule
\end{tabular*}
\end{table}

\begin{table}[t]
\caption{Sensitivity of the Proposed Method to DTX Weights}
\label{tab:sensitivity}
\centering
\footnotesize
\setlength{\tabcolsep}{2pt}
\renewcommand{\arraystretch}{1.15}
\begin{tabular*}{\columnwidth}{@{\extracolsep{\fill}}lccc@{}}
\toprule
Parameter & vs.\ Fixed plans & vs.\ Routing only & Time [s]\\
\midrule
Baseline (Table~\ref{tab:dtx_parameters}) & 19.68\% & 13.01\% & 527.25\\
$\xi_1=\xi_2=\xi_3=1/3$ & 14.63\% & 9.21\% & 531.63\\
$\xi_1=\xi_2=0.3$, $\xi_3=0.4$ & 12.18\% & 7.36\% & 514.62\\
$\xi_1=0.2$, $\xi_2=0.6$, $\xi_3=0.2$ & 36.74\% & 29.83\% & 749.11\\
\bottomrule
\end{tabular*}
\end{table}

\section{Conclusion}
\label{sec:conclusion}
This paper investigated the improvement of average trip quality in urban mixed traffic networks comprising private vehicles, autonomous vehicles, and buses under limited service capacity.  We extended the DTX to evaluate trip quality across all three travel modes and developed an integrated framework that combines DTX-based mode recommendation with dynamic routing guidance. Simulation results involving 1,500 travelers on a real-world road network in Shanghai show that the proposed framework increases mean realized DTX by $13.01\%$ over dynamic routing guidance alone and by $19.68\%$ over fixed travel modes with preplanned routes. These results demonstrate that reallocating limited autonomous vehicles and bus seats can improve overall trip quality beyond route adaptation alone, while also reducing travel times for private-vehicle users. Future work will incorporate heterogeneous traveler preferences into mode recommendations and further investigate ways to improve trip equity.

\microtypesetup{expansion=false,protrusion=false}
\bibliographystyle{IEEEtran}
\bibliography{ECC_2027}

\end{document}